# Scalable and massively parallel Monte Carlo photon transport simulations for heterogeneous computing platforms

Leiming Yu,[a] Fanny Nina-Paravecino,[a] David Kaeli,[a] Qianqian Fang[b,*]

[a]Northeastern University, Department of Electrical and Computer Engineering, Boston, Massachusetts, United States
[b]Northeastern University, Department of Bioengineering, Boston, Massachusetts, United States

**Abstract.** We present a highly scalable Monte Carlo (MC) three-dimensional photon transport simulation platform designed for heterogeneous computing systems. Through the development of a massively parallel MC algorithm using the Open Computing Language (OpenCL) framework, this research extends our existing graphics processing unit (GPU)-accelerated MC technique to a highly scalable vendor-independent heterogeneous computing environment, achieving significantly improved performance and software portability. A number of parallel computing techniques are investigated to achieve portable performance over a wide range of computing hardware. Furthermore, multiple thread-level and device-level load-balancing strat- egies are developed to obtain efficient simulations using multiple central processing units (CPUs) and GPUs. © *The Authors. Preprint published by arXiv under an arXiv.org - Non-exclusive license to distribute.*

The Monte Carlo (MC) method has been widely regarded as the gold-standard for modeling light propagation inside complex random media, such as human tissues. MC, however, suffers from low computational efficiency because a large number of photons have to be simulated to achieve the desired solution quality. Sequential MC simulations require extensive computation and long runtimes, easily taking up to several hours.[1,2] In recent years, studies on massively parallel MC algorithms have successfully reduced this computational cost down to seconds or minutes, due largely to the "embarrassingly parallelizable" nature of MC and the rapid adoption of low-cost many-core processors such as general-purpose graphics processing units (GPUs). Alerstam et al.[3] first reported a proof-of-concept using GPUs to accelerate MC in a homogeneous domain. In 2009, Fang and Boas[4] reported the first GPU-accelerated MC algorithm to model light transport inside a three-dimensional (3-D) heterogeneous domain, and released an open-source tool—Monte Carlo eXtreme (MCX).

Nearly all GPU-based MC photon transport frameworks re-ported in the literature,[3,4,5,6,7] including MCX, have been written exclusively using the CUDA programming model developed by NVIDIA.[8] Because CUDA is specifically targeted for NVIDIA GPUs, most existing GPU MC codes cannot be executed on a CPU or a high-performance GPU made by other manufacturers. In recent years, a generalized parallel computing solution—Open Computing Language (OpenCL)—has emerged.[9] OpenCL was designed to target scalability and portability in high performance computing. The OpenCL specification defines an open-standard parallel programming language for multi-core CPUs, GPUs and Field Programmable Gate Arrays (FPGAs). It specifies a set of generalized programming interfaces to efficiently utilize the computing resources of dissimilar processors.[9] A program written with the OpenCL computing model can be natively executed on cross-vendor processors, including not only NVIDIA GPUs, but also Intel and AMD CPUs and GPUs. Furthermore, the OpenCL model employs a just-in-time (JIT) compilation model for parallel code execution.[10] The OpenCL JIT compiler translates the source code, referred to as a "kernel", to device assembly at runtime. This allows processor-specific optimizations to be applied to the code, achieving improved portability and efficiency.

---

*Address correspondence to: Qianqian Fang, E-mail: q.fang@neu.edu

This work aims to improve and generalize our previously developed massively parallel photon transport simulation platform through the adoption of a heterogeneous computing framework using the OpenCL programming model. The generalized algorithm permits users to launch efficient photon transport simulations on not only NVIDIA GPUs, but also CPUs, GPUs and systems-on-a-chip processors, made by many vendors.[11]

The porting of MCX CUDA kernels to the OpenCL programming framework is relatively straightforward. A diagram of the generalized MC algorithm (MCX-CL) is shown in Fig [1]. The simulations start on the host (a CPU) by processing the user's inputs. The host then decides on how to partition the total number of simulated photons modeled based on the targeted hardware characteristics to best leverage multiple computing devices (see below). The photon simulation kernel is then dynamically compiled by the OpenCL's JIT compiler for each device. The simulation parameters, including domain settings, optical properties and independent random number seeds for each thread, are allocated, initialized and copied to each device. Once this preparation step is complete, the host instructs all activated devices to start photon transport simulations simultaneously. Each computing device launches a specific number of parallel computing threads, determined by the respective hardware settings (discussed below). Within each computing thread, a photon simulation loop (Fig. 1 in Ref. [4]) is carried out. The host waits for all devices to complete the simulation, and then reads the solutions (3-D fluence maps and detected photon data) back to the host memory. Post-processing is then performed to yield the final solution.

Several observations have been made during the implementation of MCX-CL. On a heterogeneous system, the JIT compiler and the execution library of different devices are independently implemented by their respective vendors. As a result, the same kernel may exhibit different execution behaviors on different OpenCL implementations. For example, using the AMD OpenCL implementation, multiple MCX kernels launched in the command-queue are executed asynchronously (i.e., in parallel). With the NVIDIA OpenCL library, however, kernels in the same command-queue are serialized. One has to launch multiple threads in order to use multiple NVIDIA devices in parallel. Moreover, the AMD OpenCL library supports both AMD GPUs and CPUs, but the NVIDIA OpenCL implementation only supports NVIDIA GPUs. Another caveat of OpenCL is the lack of intrinsic atomic operations for floating-point numbers. A workaround has been proposed[12] when such operations are desired.

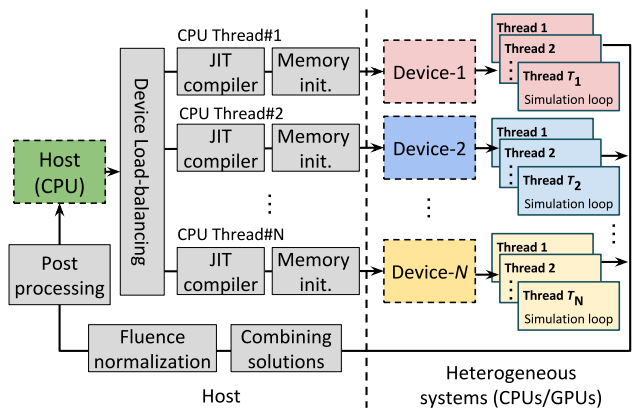

**Fig. 1** Generalized parallel Monte Carlo photon transport simulation workflow for heterogeneous systems.

Next, we characterize and optimize MCX-CL simulation performance across a range of devices, including CPUs and GPUs produced by Intel, NVIDIA and AMD. We first run a profile-based analysis on a small set of selected devices, generalize the observations and then develop a number of optimization strategies that can deliver portable performance to a wide range of devices. For profiling, we use AMD's CodeXL toolset for AMD CPUs and GPUs, and VTune Amplifier for Intel CPUs/GPUs.

A number of observations can be made from our profiling results. First, the MCX-CL kernel is compute-intensive. On an AMD R9 Nano GPU, 91 million computing instructions are executed when running our test benchmark problem (B1, see below); in comparison, only 0.5 million memory instructions were executed. Second, the number of parallel threads that MCX-CL can launch and execute is bounded by the available register space—the fastest memory in the device. For the AMD R9 Nano GPU, the available "vector register" space can only accommodate up to 768 threads (divided into 12× 64-thread groups; a 64-thread group is referred to as a "wavefront" in AMD's architecture) to run simultaneously inside a ompute unit (CU, also called a "multiprocessor" in NVIDIA literature). The third observation is that the complex workflow of the MC simulation algorithm results in 62% "thread divergence", which means that 62% of the time, only a subset of the threads inside a wavefront is executing instructions—caused by the presence of if/then/else branches. Because all 64 threads inside a wavefront are designed to execute instructions in lock-step fashion (single-instruction multiple threads), in the event that a subset of the threads need to take a different execution path, the wavefront has to be serialized, resulting in low execution efficiency.

With these key characteristics in mind, we have implemented multiple optimization strategies to maximize MCX-CL's simulation efficiency. First, to make the mathematical computation more efficient, we have utilized the "native" math functions—a set of functions with hardware-dependent accuracy provided by the OpenCL library (referred to Opt1). Second, to better utilize the available computing resources (in particular, register space), we have developed an automatic algorithm to calculate a "balanced" number of threads to ensure that all available registers are occupied (referred to Opt2). Generally speaking, a low thread number can result in low-utilization of computing resources, whereas an excessively high number of threads can result in overhead due to frequent switches between queued thread blocks. An optimized thread number can balance resource utilization to address both inefficiencies. In this work, this thread number is estimated by multiplying the maximum concurrent threads per compute unit with the available compute units on the GPU. Additionally, we simplify the control flow of the kernel (referred to Opt3), aiming to reduce thread divergence. This is also expected to reduce the complexity of the kernel, providing the JIT compiler with a better chance to optimize the execution and allocate fewer registers.

In Fig. 2, we report the MCX-CL simulation speed (in photons/ms) for 3 benchmarks (B1, B2 and B2a), before and after applying the aforementioned optimization strategies. Our baseline simulation is configured with a fixed thread number ($N$=214) and a work-group size of 64. All 3 benchmarks simulate $10^8$ photons inside a $60 \times 60 \times 60$ mm$^3$ domain, with an absorption coefficient $\mu_a = 0.005$ mm$^{-1}$, a scattering coefficient $\mu_s = 1.0$ mm$^{-1}$, an anisotropy $g = 0.01$ and a refractive index $n = 1.37$. The medium outside of the cube is assumed to be air. In B2 and B2a, a spherical inclusion ($\mu_a = 0.002$ mm$^{-1}$, $\mu_s = 5.0$ mm$^{-1}$, $g = 0.9$, $n = 1.0$) of radius 15 mm is placed at the center of the cube. In B1, a photon is terminated when it arrives at the cube's boundary; whereas in B2 and B2a, a reflection calculation is performed at the sphere and cube boundaries based on Snell's law. The difference between B2 and B2a is that B2a applies atomic operations to avoid data-races when accumulating the fluence rate in each voxel, whereas B2 uses non-atomic floating-point additions.[4] In all 3 cases, a pencil beam along $+z$-axis enters the domain at (30, 30, 0) mm. Each speed value reported is obtained by running 3 simulations and selecting the highest speed. All tests were performed on Ubuntu 14.04, using the nvidia-375 driver for NVIDIA GPUs, amdgpu-pro 16.30.3 for AMD GPUs and opencl-1.2-6.2 for Intel CPUs and GPUs. All simulations are verified to produce correct solutions. For comparison purposes, we also run the B2a benchmark using MCX (implemented in CUDA) on all NVIDIA GPUs. A speed comparison between MCX-CL and MCX is shown as an inset in Fig. 2.

From Fig. 2, the first two optimization techniques consistently produced faster simulations, although the magnitude of the improvements vary from device to device. The acceleration due to hardware-optimized math functions (Opt1) yielded some decent speed-up on AMD GPUs (7% to 12%), and Intel GPUs and CPUs (11% to 17%), and a smaller improvement on NVIDIA GPUs (3% to 10%). Combining Opt1 with Opt2 (i.e., optimized thread/workgroup size), we have observed a significant improvement for the AMD GPUs (63% to 74%), along with a moderate improvement for Intel and AMD CPUs (12% to 21%); the speed of NVIDIA GPUs is also noticeably improved (6% to 12%). However, the results when applying simplification of control flow (i.e. Opt3) are mixed—for some NVIDIA GPUs (1080Ti, 980Ti, Titan X, 1050Ti), a noticeable speedup was observed; for two other NVIDIA GPUs (1080, 590) and all AMD and Intel CPUs/GPUs, we encountered a minor reduction in speed (1% to 7%). We want to note that the GTX 1050Ti experienced a 1.9× speedup with the help of Opt3. We believe the variation in speedup when applying control flow simplification is a result of the complex interplay between kernel complexity and compiler heuristics when optimizing the kernel. Nevertheless, the advantage of using a GPU over a CPU in photon transport simulation is clear. The AMD RX Vega 64 GPU performs $23\times$ faster in B2/B2a tests compared to using the dual-Xeon E5-2658v3 CPUs (with all 48 CUs), $61\times$ faster than the i7-7700k CPU (8 CUs) and $42\times$ to $49\times$ faster than Ryzen 1700X CPU (16 CUs). Moreover, comparing the runtimes between B2 and B2a on different devices, the average overhead due to atomic operations is only 5% for all NVIDIA GPUs newer than GTX 590, which experiences a 138% overhead; the average overhead is 31% and 46% for AMD and Intel GPUs, respectively. It is interesting to note that the B2a benchmark actually runs faster (~20%) than the B2 test on the NVIDIA 980Ti and Titan X (both belong to the "Maxwell" architecture) when all three optimizations are used. We believe this is related to architecture-specific compiler optimizations.

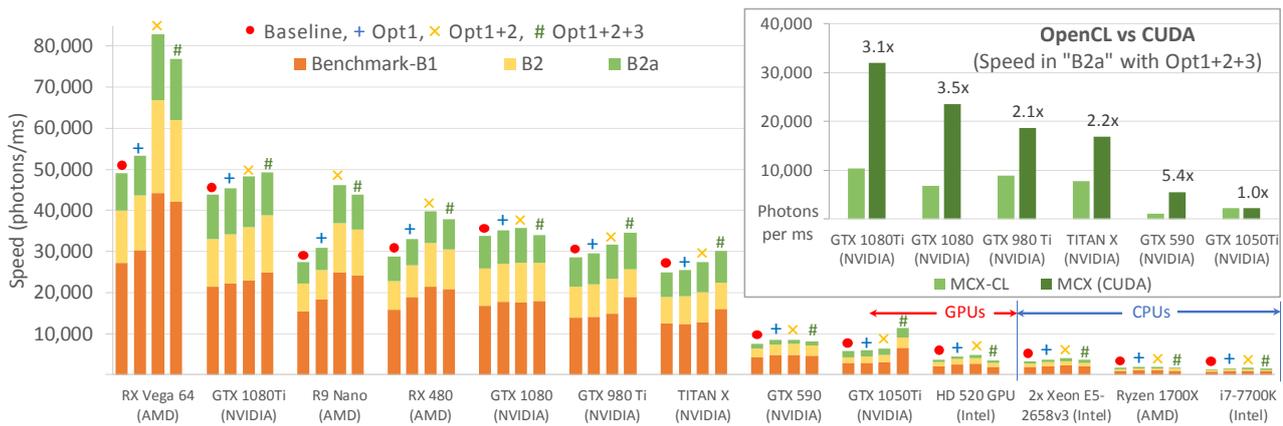

**Fig. 2** The MCX-CL simulation speed (photons/ms) on different computing devices after applying 3 optimization schemes: Opt1: using hardware-native math library; Opt2: using optimized thread configuration; Opt3: reducing thread divergence. The throughputs in the B1, B2 and B2a benchmarks are shown as a stacked-bar and the 4 bars for each hardware are Baseline (●), Opt1 (+), Opt1+2 (×), Opt1+2+3 (#), displayed from left to right. The inset shows the speed comparison between the OpenCL and CUDA versions of the algorithm on NVIDIA GPUs.

We also estimate the throughput per core (i.e. a stream processor in a GPU or a physical core in a CPU) and throughput per watt for all tested devices using the B1 benchmark with Opt1 and Opt2. Without surprise, CPUs report a significantly higher per core performance than GPUs (256 and 143 photons/ms/core for i7-7700K and Ryzen 1700X, respectively, comparing to 115 photons/ms/core for Intel HD 520 GPU and on average 9 and 6 photons/ms/core for AMD and NVIDIA GPUs, respectively). This suggests that although CPUs have more powerful cores, GPUs excel in MC simulations with many less powerful cores. For the throughput per watt calculations, we divide the throughput by the thermal design power of each processor. The Intel HD 520 GPU reports the highest power efficiency at 184 photons/ms/W, followed by AMD (145 photons/ms/W) and NVIDIA (60 photons/ms/W) GPUs; that for the CPU is between 11 to 12 photons/ms/W.

From the inset in Fig. 2, it appears that the CUDA-based MC is $2.1\times$ to $5.4\times$ faster than the OpenCL version on NVIDIA GPUs, except for the GTX 1050Ti. It is well-known that NVIDIA does not fully support OpenCL, as the current driver lacks the latest features supported by the hardware, such as floating-point atomic operations (natively supported in CUDA), therefore, resulting in the lower performance.

To efficiently run MCX-CL simulations in a heterogeneous computing environment, we have also investigated dynamic workload balancing strategies. Two types of load-balancing optimizations have been investigated: 1) improving load balance across all threads within a single execution and 2) improving load balance between computing devices (GPUs and CPUs) when multiple devices are simultaneously used.

To address the first challenge, we have developed an in-workgroup dynamic load-balancing strategy to reduce the runtime differences between different threads. In this scheme, the total photon count is first divided by the number of launched workgroups (also called a "block" in NVIDIA CUDA) as the target workload of each workgroup. Within each workgroup, each thread first checks if there are any remaining photons, if so, the thread will launch a new photon and decrease the group workload by 1; otherwise, the thread is terminated. The group workload is an integer stored in the local memory and is "atomically" decreased by each thread.

In Fig. 3(a), we show a comparison between an equal distribution of photons between threads (thread level) and the workgroup dynamic load-balanced simulations (workgroup level). On NVIDIA's GPUs, the dynamic workload generates a minor (1% on average) improvement over the uniform thread workload; on AMD GPUs, a 13% speedup is observed.

Because MCX-CL supports photon simulations with multiple computing devices, to maximize performance in such cases, an efficient device-level load-balancing strategy is needed. On most of the tested devices, the run-time ($T$) of MCX-CL exhibits a roughly linear relationship with the size of the workload (photon number $n$) as $T = a \times n + T_0$. The non-zero intercept $T_0$ is related to the host and device overhead. Both the slope $a$ and intercept $T_0$ are device-dependent. For each device, $a$ and $T_0$ can be estimated by running two pilot simulations with small photon numbers (here we use $n_1 = 10^6$ and $n_2 = 5 \times 10^6$ for such estimations).

When multiple devices are run concurrently, the "optimal" partitioning of the total workload requires us to solve a linear-programming problem. Here, three device-level load-balancing strategies are studied by distributing the total photon number using S1: the number of stream-processors (i.e., cores), S2: the throughput of the device estimated using $1/a$, and S3: the solution to a linear programming problem (using $fminimax$ in MATLAB). These strategies are compared to the "ideal" case by summing the individual device speeds.

In Fig. 3(b), we compare the simulation speed using Benchmark B1 and multiple computing devices with different capabilities. The total photon number is partitioned based on the three algorithms mentioned above. From the results, both the throughput (S2) and optimization-based (S3) load partitioning methods achieve a 10% to 14% speedup over the core-based approach (S1). Based on the near-identical results for S2 and S3, we conclude that throughput (approximated by $1/a$) can serve as a practical metric for multi-device load partitioning. For the 4 devices tested (1080Ti, 980Ti, R9 Nano, RX480), we find an overhead ($T_0$) of 53, 63, 631 and 652 ms, respectively, accounting for 1%, 0.8%, 12% and 11% of the total runtime at $n = 10^8$, respectively. A simple load-balancing scenario is tested and shown in Fig. 3(c), in which 1 to 8 identical GPUs (NVIDIA GTX 1080Ti) are simultaneously used in a single simulation. A nearly linear speedup is observed in all 3 benchmarks; in comparison, the ideal cases (assuming no overhead) are shown as dashed lines.

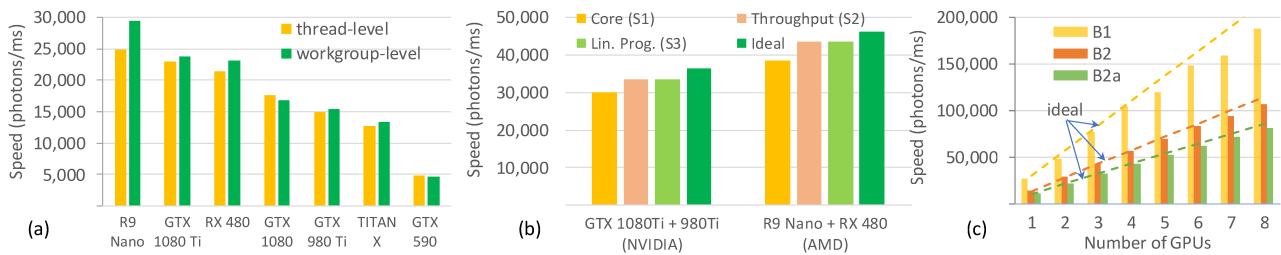

**Fig. 3** Validation of workload balancing strategies. (a) Comparison between thread- and workgroup-level load-balancing approaches using Benchmark B1, (b) comparison between 3 device-level load-balancing strategies, and (c) acceleration using 1× to 8× NVIDIA 1080Ti GPUs for the 3 benchmarks; linear acceleration (ideal case) is shown as dashed lines.

In summary, we have successfully implemented 3-D photon transport simulations using OpenCL to support a heterogeneous computing environment and multivendor hardware. Guided by profiling results, we explored various optimization techniques to improve simulation speed, and achieved a 56% average performance improvement on AMD GPUs, 20% on Intel CPUs/GPUs and 10% on NVIDIA GPUs. We also observed a significant speed gap (2.1× to 5.4×) between the CUDA-based MC simulation (MCX) and MCX-CL on most NVIDIA's GPU, reflecting the vendor's priority in supporting CUDA. We expect such underperformance will be reduced in the future as NVIDIA updates its OpenCL driver. Although the profiling analyses were only performed on selected devices, our optimization strategies show very good scalability and speed improvements on a range of tested devices, including GPUs newer than those being profiled. In addition, workgroup-level and device-level load-balancing strategies have been investigated. Our dynamic workgroup load-balancing strategy produced a 1% and 13% speedup for NVIDIA and AMD GPUs, respectively. When multiple computing devices are used concurrently for photon simulations, efficient load-partitioning strategies, based on the device throughput and linear programming models, achieved higher throughput than core-based load-partitioning.

The availability of MCX-CL makes high performance photon transport simulations readily available on a large array of modern CPUs, GPUs and FPGAs. Improved computational speed can be obtained by launching simulations on multiple computing devices, even if from different vendors. Furthermore, our insights on the Monte Carlo simulation kernel generalize our previous findings from NVIDIA GPUs to a heterogeneous computing environment. For the next step, we will implement mesh-based MC[13] for heterogeneous computing systems and compare execution performance to MCX and MCX-CL. The source code for MCX-CL is available at http://mcx.space/mcxcl/.

*Disclosures*

No conflicts of interest, financial or otherwise, are declared by the authors.

*Acknowledgments*

The authors acknowledge funding support from the National Institutes of Health (NIH) under Grant Nos. R01-GM114365 and R01-CA204443. We also thank NVIDIA for the donation of the TITAN X GPU and Dr. Enqiang Sun for his help on utilizing multiple NVIDIA GPUs.